\documentclass[9pt,a4paper]{article}
\usepackage{cite,epsfig,amsmath}
\usepackage[usenames,dvipsnames]{color}
\usepackage[pagebackref=true, colorlinks=true]{hyperref}
\definecolor{redish}{rgb}{0.7,0.2,0.0}  
\definecolor{bluish}{rgb}{0.2,0.5,0.8}
\hypersetup{linkcolor=redish,          
                  citecolor=blue,        
                  filecolor=magenta,      
                  urlcolor=bluish}          
\numberwithin{equation}{section}
\numberwithin{equation}{subsection}
\def\A{\bar A}
\def\Q{\bar Q}
\def\rhob{\bar\rho}
\def\sigmab{\bar\sigma}
\begin{document}
\title{Charged Quantum Black Holes : Thermal Stability Criterion}
\author{Abhishek Majhi\footnote{abhishek.majhi@saha.ac.in} and Parthasarathi Majumdar\footnote{parthasarathi.majumdar@saha.ac.in}  \\A P C Division \\Saha
Institute of Nuclear Physics \\Kolkata 700064, India.} 
\date{24.08.2011}
\maketitle
\begin{abstract}

A criterion of thermal stability is derived for electrically charged {\it
  quantum} black holes having large horizon area (compared to Planck area), as an inequality between the
mass of the black hole and its microcanonical entropy. The derivation is based
on key results of Loop Quantum
Gravity and  equilibrium statistical mechanics of a grand canonical ensemble,
with Gaussian fluctuations around
an equilibrium thermal configuration assumed here to be a quantum {\it isolated}
horizon. No aspect of classical black hole geometry is used to deduce the
stability criterion. Since, no particular form of the mass function is used {\it a priori}, our stability criterion provides a platform to test the thermal stability of a black hole with a given mass function. The mass functions of the two most familiar charged black hole solutions are tested as a fiducial check. We also discuss the validity of the saddle point approximation used to incorporate thermal fluctuations. Moreover, the equilibrium Hawking temperature is shown to have an additional quantum correction over the semiclassical value.

\end{abstract}


\section{Introduction}
\label{INTRO}
 
 The idea of a {\it quantum} black hole has been articulated
longer than a
 decade ago within Loop Quantum Gravity \cite{ack},
\cite{kmplb}, \cite{kmprl},
 \cite{abck}. This is an effective description
wherein a three dimensional
 $SU(2)$ Chern Simons Theory \cite{ack},
\cite{km3}, \cite{km4} coupled to point-like
 sources on punctures made by
edges of spin networks describing bulk quantum
 geometry, is shown to
describe a quantum {\it isolated} horizon \cite{firstlaw},
 \cite{ak} or, in
other words, a {\it null trapping horizon} \cite{hayw1}. The
 dimensionality
of the Hilbert space of the Chern Simons theory leads, for
 large horizon
areas (in comparison to the Planck
 area) to the Bekenstein-Hawking area law
together with universal quantum corrections
 \cite{kmprl}. These results have
been rigorously reviewed and rederived
 recently by a number of authors
{\cite{agui}, \cite{per}, \cite{sahl}. 

Quantum black holes not isolated
from an ambient thermal reservoir have also been considered in the past
\cite{dmb}, \cite{cm1,cm2,cm3}, \cite{pm1}. In
 this approach one uses
certain key results of Loop Quantum Gravity like the
 discrete spectrum of
the area operator \cite{rov,thie} and the central
 assumption that the
thermal equilibrium configuration is indeed an isolated
 horizon whose
microcanonical entropy, including quantum spacetime fluctuations
 have
already been computed via Loop Quantum Gravity. The idea here has been
 the
study of the interplay of thermal and quantum fluctuations, and a
 criterion
for thermal stability of such horizons has been obtained \cite{cm3},
\cite{pm1,pm2}, using a `thermal holographic' description involving a
canonical ensemble and incorporating Gaussian thermal fluctuations. The
generalization to horizons carrying charge has also been attempted, using a
grand canonical ensemble, even though a somewhat ad hoc mass spectrum has
been
 assumed \cite{cm2}.

In this paper, we attempt to rederive a
thermal stability criterion for charged quantum horizons, {\it without} any ad
hoc assumptions on the mass spectrum. With the benefit of hindsight, arguments
which place the earlier formulation on a more solid footing are presented,
together with novel aspects which enable us to sidestep earlier
restrictions. A comparison with semiclassical thermal stability analyses of
black holes \cite{monteiro} is made wherever possible. The range of validity of
the saddle point approximation around the equilibrium configuration is
examined to ensure the self-consistency of the Gaussian approximation.

The paper is organized as follows: In Subsection(\ref{HE}) there is a short account on the
mass associated with the horizon of a black hole. In the rest of Section(\ref{TH}) we
review the idea of Thermal Holography as laid out in \cite{pm2}; the primacy
of the boundary partition function of a Grand Canonical ensemble, in
situations where the bulk Hamiltonian is a constraint, is 
established. The boundary partition function is then evaluated in
Section(\ref{SAGTF}) within the saddle point approximation, choosing an
isolated horizon as the equilibrium configuration. Valid existence of the
saddle point is shown to lead to the thermal stability criteria in terms of a
second order partial differential inequalities involving the equilibrium mass
and the microcanonical entropy. Solving one of those inequalities yields a
more accessible form of the stability criterion in terms of a competition
between `energy-driven' and `entropy-driven' aspects of the system.  In
Section(\ref{VSPA}) the range of validity of the saddle point approximation
for the different cases is discussed. 
thermal stability criterion with semiclassical thermal stability properties of
Reissner-Nordstrom and Anti-de Sitter Reissner-Nordstrom black holes is
given. Section(\ref{DISC}) contains our concluding remarks.

\section{Thermal Holography}
\label{TH}

\subsection{Horizon Energy}
\label{HE}

For a consistent Hamiltonian evolution for spacetimes admitting internal
boundaries (isolated horizons, representing black holes at equilibrium) there
must be a first law associated with each internal boundary($b$), assumed to
be a null hypersurface with the properties of a `one-way membrane'
\cite{firstlaw,ak} given by
\[\delta E^t_{b}=\frac{\kappa^t}{8\pi}\delta A_{b} + \Phi^t \delta Q_{b}\]
where $E^t_{b}$ is the classical energy function associated with the horizon,
$\kappa^t$ and $\Phi^t$ are the surface gravity and the electric potential
respectively of the horizon, $Q_{b}$ is the horizon charge. All the quantities
are defined for a particular choice of time evolution vector field
$t^{\mu}$. The family of time evolution vector fields $[t^{\mu}]$ satisfying
such first laws on the horizon are the permissible time evolution vector
fields. These evolution vector fields also need to satisfy other boundary
conditions. Each of these time evolution vector fields associates a classical
energy function with the horizon which is  a function of area and charge for
Einstein-Maxwell theory. In arbitrary non-stationary cases (radiation may be
present arbitrarily close to the horizon) for a particular time evolution
vector field $(t)$, the Hamiltonian formulation yields 
\[H^t=E^t_{ADM}-M^t_{b}\] where $H^t$ = Hamiltonian associated with the
spacetime region in between the black hole boundary($b$) and the boundaries at
infinity, $M^t_{b}$ = the mass associated with the horizon ($b$),
$E^t_{ADM}$ = the usual ADM energy associated with the spatial boundary at
infinity for a permissible vector field $t^{\mu}$. $H^t$ is the Hamiltonian of
the covariant phase space, which is the space of various class of
solutions of the Einstein equations admitting
internal boundaries. For stationary spacetimes the global timelike Killing field
$(\xi^{\mu})$ is the time evolution vector field. On physical ground one can
say that there is nothing between the internal boundary and the boundary at
infinity for stationary spacetimes, hence $H^{\xi}=0$.  On mathematical ground one
can argue that in  the Hamiltonian framework, for the stationary black hole
solutions,  the total Hamiltonian function  $H^{\xi}$(which generates
evolution along $\xi^{\mu}$), must vanish as a first class constraint on the phase space
\cite{mih,ak}. This gives $M^{\xi}_{b}=E^{\xi}_{ADM}$. This is exactly what has been proved
in another manner in the literature : that in stationary black hole spacetimes the
ADM mass equals the energy of the black hole. Hence it is logical to identify
$E^{\xi}_{b}$ with the horizon mass $M_{b}$ in the stationary case. The difference
for an arbitrary non-stationary case is that $H^t\neq 0$. Thus it can be called as the
mass associated with the Isolated Horizon in an active sense, that can change from one dynamic equilibrium
situation to another satisfying the first law. Here, one should be careful
that this mass associated with the isolated horizon is completely
physical and is not to be confused with the Hamiltonian of the $SU(2)$
Chern-Simons theory which describes horizon field degrees of freedom. The Hamiltonian
for Chern Simons theory vanishes identically, since the theory is topological
and insensitive to small metric deformations.

Clearly, the horizon mass is {\it not} affected by boundary conditions at
asymptopia. It is defined {\it locally} on the horizon without referring to the
asymptotic structure at all. The asymptotic conditions only modify
the energy associated with the boundary at infinity and the bulk equation of
motion (Einstein equations) \cite{ak,miha}. The Hamiltonian framework
discussed above is equally applicable for asymptotically flat and AdS
spacetimes.

\subsection{Quantum Geometry}
\label{QG}

For a classical spacetime with boundary, boundary conditions determine the
boundary degrees of freedom and their dynamics. For a quantum spacetime, on
the other hand, fluctuations of the boundary degrees of freedom have a `life'
of their own (see for instance ref.\cite{ack}). Consequently, the Hilbert
space of a quantum spacetime with boundary has the tensor product structure
${\cal H} ={\cal H}_v \otimes {\cal H}_b$, with the subscript $v~(b)$ denoting
the bulk (boundary) component. Thus, any generic state in quantum geometry,
$|\Psi\rangle$, admits the expansion
\begin{eqnarray}
|\Psi \rangle = \sum_{v,b}~C_{vb}~|\psi_v\rangle \otimes |\chi_b \rangle ~.
~\label{expan}
\end{eqnarray}      

In presence of electromagnetic fields, one can consider $|\psi_v \rangle$
(resp. $|\chi_b\rangle$) to be the composite quantum gravity + quantum electrodynamics
bulk (resp. boundary) state. The bulk states are annihilated by the {\it full}
bulk Hamiltonian : ${\hat H}_v |\psi_v \rangle \equiv [{\hat
 H}_{g,v} +
{\hat H}_{e,v}] |\psi_v \rangle = 0$; this is the quantum version of the
classical Hamiltonian constraint \cite{thie}. The total Hamiltonian operator
acting on the generic state $|\Psi \rangle$ has the form
\begin{eqnarray}
{\hat H_T} |\Psi\rangle = ({\hat H}_v \otimes I_b + I_v \otimes {\hat H}_b ) |
\Psi \rangle  ~\label{hamilc}
\end{eqnarray} 
 where, $I_v (I_b)$ corresponds to the identity operator on ${\cal H}_v ({\cal H}_b ) $. 

While defining the grand canonical partition function, the charge operator
($\hat{Q}$) for the black hole is also needed. It can be written in a similar
fashion like the Hamiltonian as
\[\hat{Q}|\Psi\rangle=(\hat{Q}_{v}\otimes \hat{I}_b + \hat{I}_v \otimes \hat{Q}_{b})|\Psi\rangle\] 
where $\hat{Q}_{v}$ and $\hat{Q}_{b}$ are corresponding charge operators for
the bulk states $|\psi_v\rangle$ and the boundary states $|\chi_b\rangle$ ,
respectively. In the classical theory the charge of a black hole is defined on
the horizon i.e the internal boundary of the four dimensional spacetime
(e.g. one can see how charge can be properly defined for spacetimes admitting
internal boundaries in Einstein-Maxwell or Einstein-Yang-Mills theories in
\cite{firstlaw}). There is \textit{no} charge associated with the bulk  black
hole spacetime, i.e. $Q_v \approx 0$ , which is basically the Gauss law
constraint for electrodynamics. Hence, its quantum version is \textit{assumed}
to be \[\hat{Q}_v|\psi_v\rangle = 0. \] 

Combining the quantum constraints on the Hamiltonian and charge operators we
can define a new quantum constraint as
\[\hat{H}^{\prime}_v|\psi_v\rangle=0\]
 where $\hat{H}^{\prime}_v\equiv
\hat{H}_{Tv}-\Phi\hat{Q}_v$ and $\Phi$ may be any function. But in our case it
is a physically significant quantity which will  be defined in the next
paragraph. The implications of these quantum constraints will be seen during
the construction of the grand canonical partition function.

\subsection{The Partition Function}
\label{PF}

Let us consider a grand canonical ensemble of massive charged black holes
immersed in a heat bath at some finite temperature with which it can exchange
energy and charge as well. We construct a partition function
for the thermodynamic system. Using the usual definition of the grand
canonical partition function we write
\[Z_G~=~Tr~\exp-\beta\hat{H}_T+\beta\Phi\hat{Q}\] where the trace is taken
over all states. $\Phi$ is the electrostatic potential and $\hat{Q}$ is the
charge operator for the black hole. To write it in the explicit form first we
write down a general quantum state of the black hole as follows
\[|\Psi\rangle~=~\sum_{v,b} c_{vb}~|\psi_v\rangle\otimes|\chi_b\rangle\]
Now, we can write the partition function as 
\begin{eqnarray}
Z_{G}&=& \sum_{v,b} |c_{vb}|^2\langle \chi_b|\otimes\langle \psi_v|\exp -\beta\hat{H}_T  + \beta\Phi\hat{Q}|\psi_v\rangle\otimes|\chi_b\rangle \nonumber\\
&=& \sum_{v,b} |c_{vb}|^2\langle \chi_b|\otimes\langle \psi_v|\exp -\beta\hat{H}^{\prime}|\psi_v\rangle\otimes|\chi_b\rangle \label{parti}
\end{eqnarray}
where, $\hat{H}^{\prime}=\hat{H}_T-\Phi\hat{Q}$. Writing the new operator $H^{\prime}$ as $(\hat{H}^{\prime}_v\otimes \hat{I}_b + \hat{I}_v \otimes \hat{H}^{\prime}_b)$ and using $\hat{H}^{\prime}_v|\psi_v\rangle= 0$, the
partition function comes out to be equal to the boundary partition function
only i.e. $Z_G = Z_{Gb}$, where $Z_{Gb}$ is the boundary partition function
for the charged isolated horizon, given by
\begin{eqnarray}
Z_{Gb} = Tr_{b}\exp-\beta(\hat{H}_b - \Phi\hat{Q}_b)\nonumber  
\end{eqnarray}
where it is assumed that the boundary states can be normalized through the
squared norm $\sum_v|c_{vb}|^2\langle\psi_v|\psi_v\rangle=|C_b|^2$. This is
analogous to the canonical ensemble scenario described in \cite{pm1}. 

Now, the spectrum of the boundary Hamiltonian operator is still unknown in
Loop Quantum Gravity. So we \textit{assume} that the spectrum of the boundary
Hamiltonian operator is a function of the discrete area spectrum and the
charge spectrum of the area and the charge operators associated with the
horizon, respectively\footnote{Actually this second assumption follows from
the discussion in Subsection(\ref{HE}) \cite{firstlaw,ak} for spacetimes
admitting weakly isolated horizons where there exists a mass function
determined by the area and charge associated with the horizon. This is an
extension of that assumption to the quantum domain.}. The charge spectrum is
equispaced on general physics grounds of charge quantization. The area
spectrum, in Loop Quantum Gravity, can be {\it approximately} taken to be equispaced for large
area black holes.
For large area black holes, it is a
well known fact in literature \cite{agpm} that the major contribution to
the entropy comes from the lowermost spins. Of course the higher
spins contribute, but their contribution is exponentially
suppressed. Thus, the effect of the corrections due to the inclusion
of these higher spins is much more a technical aspect rather than
physical. Hence, only spin $1/2$ contributions for all punctures is
taken into account, which leads to the equispaced area spectrum as
an {\it approximation}, linear in the number of punctures .

In a basis in which both area and charge operators are diagonal, the partition
function can be written as
\begin{eqnarray}
Z_G~=~\sum_{k,l} g(k,l)\exp -\beta\left[E(A_k,Q_l)-\Phi Q_l\right] 
\end{eqnarray} 
 where $g(k,l)$ is the degeneracy corresponding to the area
eigenvalue $A_k$ and the charge eigenvalue $Q_l$. $k,l$ are the area and
charge quantum numbers respectively. As we are interested in  regime of the
large area and charge eigenvalues $(k\gg1, l\gg1)$, Poisson resummation formula
is applied \cite{cm2} to approximate the summation to an integration given by
\begin{eqnarray}
Z_{G}~=~\int dx~dy~\exp-\beta\left\{E\left[A(x),Q(y)\right] -
\Phi Q(y)\right\}~g\left[A(x),Q(y)\right]
\end{eqnarray}
where $x$ and $y$ are area and charge quantum numbers in the
continuum limit of $k$ and $l$ respectively. Since $A=A(x)$ and $Q=Q(y)$, we
can write $dx=\frac{dA}{A_x}$ and $dy=\frac{dQ}{Q_y}$ to write the partition
function in terms of area and charge as free variables as follows 
\begin{eqnarray}
Z_G &=&\int{\frac{dA~dQ}{A_x~Q_y}~g(A,Q)~ \exp -\beta\left\{E(A,Q)-\Phi
Q\right\}} \nonumber \\
&\approx& \int{dA~dQ~e^{S(A)- \beta E(A,Q)+ \beta \Phi Q}} \label{eq:zg} 
\end{eqnarray} 
where $\frac{g(A,Q)}{A_x~Q_y}=e^{S(A)}$ \cite{landau}. $S(A)$ being the
microcanonical entropy is a function of black hole area alone, as has been
established within loop quantum gravity \cite{ack,kmprl,abck}. 

\section{Stability Against Gaussian Thermal Fluctuations}
\label{SAGTF}

\subsection{Saddle Point Approximation(S.P.A.)}
\label{SPA}
Having a well defined partition function, we investigate its finiteness under
Gaussian thermal fluctuations about stable equilibrium configurations of the
black hole given by the saddle points $\left\{\A,\Q\right\}$.  Taylor expanding $(S(A)- \beta
 E(A,Q)+ \beta \Phi Q)$ about a saddle point $(\A,\Q)$ and applying the saddle point conditions one can rewrite the partition function as
\begin{eqnarray}
Z_{G}&=& ~e^{\left[S(\A)-\beta M(\A,\Q)+\beta \Phi \Q \right]} \nonumber\\
&&\times\int{e^{ -\frac{1}{2}\left[-\left\{S_{AA}(\A)-\beta M_{AA}(\A,\Q)\right\}a^ 2 + \beta M_{QQ}(\A,\Q)q^2+ 2\beta
M_{AQ}(\A,\Q)aq\right]}}~~da~dq~~~~~~~ \label{eq:parti}
\end{eqnarray}
where, we have used $M(\A,\Q)$ to indicate the equilibrium isolated horizon mass as a
function of the area and charge. The saddle point conditions imply that the
coefficients of $~a=(A- {\bar A})$ and $q=(Q-{\bar Q})$ must vanish, which yield
\begin{eqnarray}
\beta(\A,\Q) =\frac{S_A(\A)}{M_A(\A,\Q)}~,~\Phi(\A,\Q) =M_Q(\A,\Q) \label{betaphi}\label{spaconditions}
\end{eqnarray}

\subsection{Quantum Surface Gravity}
\label{QSG}

An interesting result of this statistical mechanical approach is that, it
gives rise to a quantum correction to the surface gravity which is a direct
consequence of the logarithmic corrections of the microcanonical entropy
$S=\frac{\A}{4}-\frac{3}{2}\log\frac{\A}{4}$ from loop quantum gravity. If one
calculates $\beta$ from the saddle point conditions and use it to find the
quantum surface gravity $(\kappa_{quantum})$ in terms of classical surface
gravity $(\kappa_{classical})$, there appear additional correction terms. One
does this by calculating $\beta$ from (\ref{betaphi}) and then using it in the
expression
 $\kappa_{quantum}=\frac{2\pi}{\beta}$(more appropriately, this
$\beta$ can be replaced by $\beta_{quantum}$ ). The quantum corrections to the
classical surface gravity is found to be 
\begin{eqnarray}
\kappa_{quantum}\approx\kappa_{classical} \left(1+\frac{6}{\A}\right)\label{kappacorrection}
\end{eqnarray} 
where higher order terms are neglected for large black holes $(\A>>1)$, $\A$ being the area of the weakly isolated horizon in Planck units.  One can easily check the formula by applying it to the Reissner-Nordstrom and AdS Reissner-Nordstrom cases. Since the formulation is not dependent on any particular situation (symmetry, etc.), it will also be valid for other massive charged black holes also.

\subsection{Stability Criteria}
\label{SC}
For the integral (\ref{eq:parti}) to be convergent, the Hessian matrix $H$, given by (\ref{eq:hess}) has to be \textit{positive definite} where
\begin{eqnarray}
H=
 \begin{pmatrix}
\beta M_{AA}(\A,\Q) - S_{AA}(\A)      &        \beta M_{AQ}(\A,\Q) \\
\beta M_{AQ}(\A,\Q)       &          \beta M_{QQ}(\A,\Q) 
\end{pmatrix}
\label{eq:hess}
\end{eqnarray}

The necessary and sufficient condition for the \textit{real symmetric square matrix} $H$ to be positive definite can be stated as follows \cite{matrix} - \textit{`Determinants of all the principal submatrices, including the determinant, of $H$ are positive.'} It is also a crucial point that the inverse temperature $ \beta $ must be positive. 
Hence, the necessary and sufficient conditions for the positive definiteness of the Hessian matrix lead to the following stability criteria 
\begin{eqnarray}
&& \beta ~\equiv~ \frac{S_A(\A)}{M_A(\A,\Q)} >0\label{eq:stability0}\\
&& \beta M_{AA}(\A,\Q) - S_{AA}(\A)>0\label{eq:stability1}\\
&& \det H ~\equiv~ \left\{\beta M_{AA}(\A,\Q) -
S_{AA}(\A)\right\}\beta M_{QQ}(\A,\Q)-\beta^2
M^2_{AQ}(\A,\Q)>0 \label{eq:stability2}
\end{eqnarray}

It should be noted that, for $ M_{QQ}(\A,\Q)>0$, it will suffice to check only conditions (\ref{eq:stability0}) and (\ref{eq:stability2}) (which will be the case for RN and AdS RN black holes).

\par
Here, one may wonder that how these stability criteria (\ref{eq:stability0})--(\ref{eq:stability2}) are related to the convexity property of the entropy function, which is the usual notion for thermodynamic stability. It is true that the usual notion of thermodynamic stability is related to the convexity property of the entropy function. It is also true that this convexity property of the entropy function follows from the requirement of the convergence of the partition function under Gaussian thermal fluctuations\cite{cm2,monteiro,landau}. Our stability criterion, likewise, follows from equations (\ref{eq:stability0})--(\ref{eq:stability2}) which constitute the necessary and sufficient conditions for the grand canonical partition function to be well defined. Physically, the conditions to be satisfied by the entropy Hessian in \cite{cm2},\cite{monteiro} and \cite{landau} (which imply convexity of entropy function) and the conditions (\ref{eq:stability0})--(\ref{eq:stability2}) lead to the same conclusion i.e. the finiteness of the partition function under Gaussian thermal fluctuations. In fact, one can check that the apparent dissimilarity in the mathematical structure of the Hessian in \cite{cm2} and eq. (\ref{eq:hess}) is just a manifestation of the different variables which are summed over\footnote{To understand the motivation behind this choice of variables one can see Subsection (\ref{HE}) where a detailed explanation is given.}. The equivalence between our conditions and the convexity of the entropy function is obvious for the charge-less case i.e. canonical ensemble, discussed in details in \cite{pm1}.

\subsection{The Product Ansatz}
The stability criterion (\ref{eq:stability2}) derived from the positive definiteness of the Hessian matrix, takes the form
\begin{eqnarray}
\frac{M_{AA}}{M_A} - \frac{M_{QA}^2}{M_{QQ}M_A} >\frac{S_{AA}}{S_A} \label{stcri}
\end{eqnarray}

This is a second order partial differential inequality; one can resort to solving it for $M(A,Q)$ in terms of $S(A)$ and $Q$ by trying to separate variables as one does for second order partial differential equations. Accordingly, we decompose $M(A,Q) = \mu(A) \cdot \chi(Q)$ with $\dim \mu = \dim M$ and $\dim \chi =0$. Further, $\chi(0) =1$,
so as to agree with the result in the chargeless (Schwarzschild) case. The inequality (\ref{stcri}) then reduces to 
\begin{eqnarray}
\frac{\mu}{\mu_A}~\left[\frac{\mu_{AA}}{\mu_A} - \frac{S_{AA}}{S_A} \right] >\frac{\chi_Q}{\chi} \frac{\chi_Q}{\chi_{QQ}} \label{sepcr}
\end{eqnarray}

Since the left and right sides of inequality (\ref{sepcr})  are functions of two completely independent variables $\A$ and $\Q$, for the inequality to make sense, one may set the right side (say) to a constant $1/\kappa$ say. One then obtains  
\begin{eqnarray}
\chi(\Q) = (1 + C\Q)^{\frac{1}{\kappa-1}}  \label{chiq}
\end{eqnarray}
where, $C$ is a constant with $\dim C = \dim Q^{-1}$. Substituting this in the inequality (\ref{sepcr}), we get
\begin{eqnarray}
\mu(\A) > (\alpha S + \gamma)^{\frac{\kappa}{\kappa-1}} \label{mua}
\end{eqnarray}
where, $\kappa > 1$. Setting $\gamma =0$ for simplicity, and choosing $(k_B\alpha)^{\frac{\kappa}{\kappa-1}}=M_P$, we get 
\begin{eqnarray}
\frac{M}{M_P} &>& \frac{S}{k_B} \left[ \frac{S}{k_B(1+C\Q)}\right]^{\frac{1}{\kappa-1}}  \nonumber \\
&=& \frac{S}{k_B} \left[ 1 + \frac{1}{\kappa-1} \ln \left( \frac{S}{k_B(1+C\Q)}\right)+ \cdots \right] \label{bound}
\end{eqnarray}
where the second line follows from the first in the limit $\kappa \gg 1$. It is easy to see that this bound is consistent with the chargeless case. It has to be seen how this bound checks out when the classical mass spectrum as a function of the area and charge of RN and AdS-RN black holes are used. The mass bound (\ref{bound}) is rewritten as
\begin{eqnarray}
M(\A,\Q) > S(\A) \left[ \frac{S(\A)}{1 + ( \frac{4\pi \Q^2}{l_C^2})^{1/2} } \right]^{1 / \kappa - 1} \label{mbd}
\end{eqnarray}
where we have replaced the constant $C$ in terms of $l_C$. The classical RN is given by the metric ($G=k_B=c=\hbar= 1$) 
\begin{eqnarray}
ds^2=-(1-\frac{2M}{r}+\frac{Q^2}{r^2})dt^2+(1-\frac{2M}{r}+\frac{Q^2}{r^2})^{-1}dr^2+r^2d\Omega^2\label{rn}
\end{eqnarray}
This leads to the mass associated with the RN horizon as a function of the horizon area and charge:
\begin{eqnarray}
M_{RN}(A,Q) = \left( \frac{A}{4\pi} \right)^{1/2} ~\left[ 1 + \frac{4\pi Q^2}{A} \right] \label{ma}
\end{eqnarray} 
Clearly, the charge is bounded by above according to $4\pi Q^2 \leq A$. Inserting this in equation (\ref{mbd}) leads to
\begin{eqnarray}
M(\A) > S \left[ \frac {S} {1 + S^{1/2}}\right]^{1/\kappa-1} \label{bou}
\end{eqnarray}
where, we have used $S(A) = A/4 l_P^2$ and set $l_C = 2 l_P$ for simplicity. We have reverted to the use of the Planck length $l_P = (G \hbar /c^3)^{1/2}$ as a fundamental length scale signifying onset of LQG. For black holes of very large horizon area $A >> l_P^2$, the above bound reduces to 
\begin{eqnarray}
M(\A) > S^{\frac{2\kappa - 1} {2\kappa -2}} \label{bd}
\end{eqnarray}

In the same limit the classical $M_{RN}$ obeys the upper bound $M_{RN}< 2 S^{1/2}$ which obviously violates the lower bound (\ref{bd}).

A similar analysis for the AdS-RN shows that for black holes of very large area compared to the cosmic area $A_{\Lambda} \equiv 4\pi \left(-\frac{3}{\Lambda}\right)= 4\pi l^2$, the classical black hole mass obeys the bound (\ref{bd}). This observation comes from the fact that the metric for AdS-RN is 
\begin{eqnarray}
ds^2=-(1-\frac{2M}{r}+\frac{Q^2}{r^2}-\frac{\Lambda}{3}r^2)dt^2+(1-\frac{2M}{r}+\frac{Q^2}{r^2}-\frac{\Lambda}{3}r^2)^{-1}dr^2+r^2d\Omega^2\label{adsrn}
\end{eqnarray}
where, $\frac{\Lambda}{3}=-\frac{1}{l^2}$. The mass associated with the AdS RN horizon is given by
\begin{eqnarray}
M_{ADSRN}(A,Q) = \left( \frac{A}{4\pi} \right)^{1/2} ~\left[ 1 + \frac{4\pi Q^2}{A} + \frac{A}{4\pi l^2} \right] ~\label{ma1}
\end{eqnarray} 
It is obvious that for $A >> A_{\Lambda}$, the last term in (\ref{ma1}) dominates over the other two, and in this case, the bound is indeed obeyed. In fact the charge term makes a small contribution in any event, being bounded from above by 1.

\par
\textbf{First Law :} The mass formula written in the form $M(A,Q)=\mu(A)\chi(Q)$ satisfies the first law. Differentiating $M$ we have
\[dM = \mu_A \chi dA + \mu \chi_Q dQ\]
Now, from (\ref{spaconditions}) we have $\Phi = M_Q = \mu\chi_Q$. Hence
\[dM = \mu_A \chi dA + \Phi dQ\]
Again, finding $\beta$ from (\ref{spaconditions}) and using $\kappa_{sg}=\frac{2\pi}{\beta}$ one finds that the surface gravity comes out to be $\kappa_{sg}=8\pi\mu_A\chi$(`sg' stands for surface gravity). Then, it is easy to see that this $\kappa_{sg}$ proves that the product ansatz satisfies the first law
\[dM = \frac{\kappa_{sg}}{8\pi}dA+\Phi dQ\]
Thus, from the thermodynamic point of view, this kind of solution for mass is acceptable.

\subsection{The Classical Metrics}\label{CMTS}

\textbf{Reissner-Nordstrom black hole :} In this paragraph we investigate the stability of massive charged Reissner-Nordstrom black holes against Gaussian thermal fluctuations. The classical Reissner-Nordstrom metric which is given by (\ref{rn}). The mass of the black hole in terms of area and charge as independent variables and it is given by
\begin{eqnarray}
M=\frac{1}{2}\left(\frac{A}{4\pi}\right)^{\frac{1}{2}}\left(1+\rho\right)\label{eq:RNmass}
\end{eqnarray}
where $\rho = \frac{4\pi Q^2}{A}$ is a dimensionless parameter in the chosen units. The microcanonical entropy including the logarithmic correction term from loop quantum gravity is given by
\begin{eqnarray}
S=\frac{A}{4}-\frac{3}{2}\log\frac{A}{4}\label{eq:SMC}
\end{eqnarray}
Using (\ref{eq:RNmass}) and (\ref{eq:SMC}) one finds that
\begin{eqnarray}
 && \beta =2\sqrt{\pi}A^{\frac{1}{2}}\frac{(1-\frac{6}{\A})}{(1-\rhob)}\label{rnbeta}\\
 && det H = -\frac{\pi}{2\A}\frac{(1-\frac{6}{\A})(1+\frac{6}{\A})}{(1-\rhob)}\label{rnH}
\end{eqnarray}  
where $\rhob = \frac{4\pi \Q^2}{\A}$. Conditions (\ref{eq:stability0}) and (\ref{rnbeta}) together imply $\rhob<1$, whereas,  (\ref{eq:stability1}) and (\ref{rnH}) imply $\rhob>1$. From  this contradiction one can conclude that the Reissner-Nordstrom black hole is \textit{locally unstable} against Gaussian thermal fluctuations.\footnote{This result is in agreement with what has been told in \cite{gm} i.e. \textit{an asymptotically flat, non-extremal black hole can never achieve a state of thermal equilibrium}. This paper also uses a statistical mechanical formulation without any classical metric but do not have a sound mathematical basis as a justification of the quantum statistical formulation which has been presented in our work.}

\par
\textbf{AdS Reissner-Nordstrom black hole :} In this paragraph we investigate the stability of massive charged AdS Reissner-Nordstrom black holes against thermal fluctuations. The classical AdS Reissner-Nordstrom metric which is given by (\ref{adsrn}). The mass of the AdS Reissner-Nordstrom black hole in terms of area and charge as the independent variables is given by
\begin{eqnarray}
M=\frac{1}{2}\left(\frac{A}{4\pi}\right)^{\frac{1}{2}}\left(1+\rho+\sigma\right)\label{eq:ARNmass}
\end{eqnarray}
where we have set $\frac{\Lambda}{3}=-\frac{1}{l^2}$ as $\Lambda$ is negative for AdS spacetimes and introduced the parameter $\sigma =\frac{A}{4\pi l^2}=\frac{A}{A_{\Lambda}}$. Here also the entropy is given by (\ref{eq:SMC}). Calculating the required quantities for AdS Reissner-Nordstrom case from (\ref{eq:SMC}) and (\ref{eq:ARNmass}) one finds that
\begin{eqnarray}
&& \beta =2\sqrt{\pi}\A^{\frac{1}{2}}\frac{(1-\frac{6}{\A})}{(1-\rhob +3\sigmab)}\nonumber\\
&& det H = \frac{\pi}{2\A}\frac{(1-\frac{6}{\A})(1+\frac{6}{\A})}{(1-\rhob+3\sigmab)^2}\left[\rhob-1+3\sigmab\left(1-\frac{18}{\A}\right)\left(1+\frac{6}{\A}\right)^{-1}\right]\nonumber
\end{eqnarray}
where $\sigmab =\frac{\A}{4\pi l^2}=\frac{\A}{A_{\Lambda}}$. For the conditions (\ref{eq:stability0}) and  (\ref{eq:stability2}) to be satisfied simultaneously  the parameters have the following bound given by
\begin{eqnarray}
1-3\sigmab\left(1-\frac{18}{\A}\right)\left(1+\frac{6}{\A}\right)^{-1}<\rhob<1+3\sigmab\nonumber
\end{eqnarray}
which in the  first order approximation reduces to the bound
\begin{eqnarray}
1-3\sigmab\left(1-\frac{24}{\A}\right)<\rhob<1+3\sigmab\nonumber
\end{eqnarray}
Hence, AdS Reissner-Nordstrom black holes are \textit{locally stable} against thermal fluctuations for the above range of parameters.

\section{Validity of S.P.A.}
\label{VSPA}
 To ensure the validity of the saddle point approximation, the relative r.m.s. fluctuations about the saddle point are to be checked i.e. $\Delta A_{rms}/\A=(\Delta A^2)^{1/2}/\A,\Delta Q_{rms}/\Q=(\Delta Q^2)^{1/2}/\Q$ where
\begin{eqnarray}
 \Delta A^2 = (H^{-1})_{11}= \frac{\beta M_{QQ}}{det H}~~~,~~~~~~~~\Delta Q^2
 = (H^{-1})_{22} = \frac{\beta M_{AA}- S_{AA}}{det H}\nonumber
\end{eqnarray}

\subsection{The Product Ansatz}
Using the product ansatz the mean square fluctuation of the charge comes out to be 
\begin{equation}
\Delta Q^2=\frac{(S_A\mu_{AA}-S_{AA}\mu_A)\mu_A\chi^2}{S_A(S_A\mu_{AA}-S_{AA}\mu_A)\mu\chi_{QQ}\chi-S^2_A\mu^2_A\chi^2_Q}\label{a}
\end{equation}
Now, the following quantities are calculated
\begin{eqnarray}
&&1+CQ = \left(\frac{\kappa - 1}{\mu C}\right)^{\frac{\kappa - 1}{2 - \kappa}}\Phi^{\frac{\kappa - 1}{2 - \kappa}}\nonumber\\
&&\chi=(1+CQ)^{\frac{1}{1-\kappa}}~=\left(\frac{\kappa-1}{\mu C}\right)^{\frac{1}{2-\kappa}}\Phi^{\frac{1}{2-\kappa}}\nonumber\\
&&\chi_Q=\frac{C}{\kappa-1}(1+CQ)^{\frac{2-\kappa}{\kappa-1}}~=\frac{\Phi}{\mu}\nonumber\\
&&\chi_{QQ}=\frac{C^2(2-\kappa)}{(\kappa-1)^2}(1+CQ)^{\frac{3-2\kappa}{\kappa-1}}~=\frac{C^2(2-\kappa)}{(\kappa-1)^2}\left(\frac{\kappa
    - 1}{\mu C}\right)^{\frac{3-2\kappa}{2 - \kappa}}\Phi^{\frac{3-2\kappa}{2
    - \kappa}} 
\end{eqnarray}
Using all these quantities in (\ref{a}) one finds that
\begin{eqnarray}
\Delta Q^2=\frac{\eta/S_A}{\eta\mu(2-\kappa)-S_A \mu_A^2} \label{qsquare}
\end{eqnarray}
where $\eta=S_A\mu_{AA}-S_{AA}\mu_A$. Now, condition (\ref{eq:stability2})
implies $M_{QQ}(\beta M_{AA}-S_{AA})>\beta M^2_{AQ}$. Since the R.H.S. is
positive definite $(\beta>0)$, the L.H.S. has to be positive definite. This
implies either $M_{QQ}>0~,~(\beta M_{AA}-S_{AA})>0$ or $M_{QQ}<0~,~(\beta M_{AA}-S_{AA})<0$. 
The second one is not possible as it makes $\Delta Q^2$ negative. Hence the only possibility is the first one. Since $S_A>0$, hence $(\beta M_{AA}-S_{AA})>0$ implies $\eta>0$. Keeping all these conditions in mind if one goes back to the expression (\ref{qsquare}), one finds that for $\Delta Q^2>0$, $\kappa$ must satisfy the condition $\kappa<2$.The relative r.m.s. fluctuation of the charge goes as
\begin{eqnarray}
\frac{(\Delta Q)_{rms}}{\Q}=\frac{4C\pi^{\frac{1}{4}}\left(1+\frac{12}{\A}\right)^{\frac{1}{2}}}{\left\{\left(1+\frac{6}{\A}\right)(2-\kappa)+\left(1-\frac{6}{A}\right)\right\}^\frac{1}{2}\left\{\left(\frac{\kappa-1}{\mu C}\Phi\right)^{\frac{\kappa-1}{2-\kappa}}-1\right\}}\A^{-\frac{1}{4}}\label{pacf}
\end{eqnarray}
where $\mu$ and its derivatives are calculated from the relation $A=8\pi\mu^2$. The arguement is as follows.The irreducible mass is a function of area only and it is the Schwarzschild mass. Hence, for $Q=0(\chi(0)=1)$ we have $A=8\pi\mu^2$. The mean square of area fluctuations comes out to be
\[\Delta A^2=\frac{\mu\mu_A}{\eta\mu-\frac{S_A\mu^2_A}{2-\kappa}}\] 
 Thus the relative rms fluctuation of area comes out to be
\begin{eqnarray}
\frac{(\Delta A)_{rms}}{\A}=\left(\frac{\mu\mu_A}{\eta\mu-\frac{S_A\mu^2_A}{2-\kappa}}\right)^{\frac{1}{2}}\frac{1}{16\pi\mu^2}\label{asquare}
\end{eqnarray} 
To see the variation of (\ref{asquare}) with $A$ it is written explicitly in terms of $A$ as follows
\begin{eqnarray}
\frac{(\Delta A)_{rms}}{\A}=\frac{2\sqrt{2}}{\left\{\left(1-\frac{6}{\A}\right)(2-\kappa)^{-1}-\left(1+\frac{6}{\A}\right)\right\}^{\frac{1}{2}}}\A^{-\frac{1}{2}}\label{paaf}
\end{eqnarray}
 This again shows that the constant $\kappa$ has an upper bound which is not much larger than unity. Moreover, calculation yields $\eta=-\frac{1}{64\sqrt{\pi}}(1+\frac{6}{\A})\A^{-\frac{3}{2}}$ which is a negative definite quantity(contradicts $\eta>0$). For $\eta<0$ we must have $\chi_{QQ}<0$ which in turn implies $\kappa>2$. But this step i.e. $\eta<0$ will lead to negative $\Delta Q^2$, which again implies instability. Hence $\kappa$ has to be bounded from above by 2. Considering the appropriate bounds of $\kappa$ it is seen from (\ref{pacf}) and (\ref{paaf}) that the relative r.m.s. fluctuations fall off for large area of the weakly isolated horizon i.e. $\A\gg1$ in Planck units.


\subsection{The Classical Metrics}

\textbf{Reissner-Nordstrom black hole :} The mean square fluctuations of area
and charge in this case  come out to be  negative due to the presence of the
determinant of the Hessian matrix. This emphasizes the instability of
Reissner-Nordstrom black holes against Gaussian thermal fluctuations.

\textbf{AdS Reissner-Nordstrom black hole :} The relative r.m.s. fluctuations
of area and charge in this case are found to be 
\begin{eqnarray}
&&\frac{\Delta A_{rms}}{\A}=\left[\frac{8(1-\rhob+3\sigmab)}{\left(1+\frac{6}{\A}\right)\left\{\rho-1+3\sigmab\left(1-\frac{18}{\A}\right)\left(1+\frac{6}{\A}\right)^{-1}\right\}}\right]^{\frac{1}{2}}\A^{-\frac{1}{2}}\label{adsrnaf}\\
&&\frac{\Delta Q_{rms}}{\Q}=\frac{\sqrt 3}{\Phi^2}\left[\frac{(1-\rhob+3\sigmab)\left(1-\frac{2}{\A}\right)\left\{\rhob+\sigmab\left(1-\frac{18}{\A}\right)\left(1-\frac{2}{\A}\right)^{-1}-\frac{1}{3}\left(1+\frac{6}{\A}\right)\left(1-\frac{2}{\A}\right)^{-1}\right\}}{\left(1-\frac{6}{\A}\right)\left(1+\frac{6}{\A}\right)\left\{\rho-1+3\sigma\left(1-\frac{18}{\A}\right)\left(1+\frac{6}{\A}\right)^{-1}\right\}}\right]^{\frac{1}{2}}\A^{-\frac{1}{2}}\nonumber\\
~~~\label{adsrncf}
\end{eqnarray} 
where the equilibrium charge has been replaced in terms
of the electric potential $\Phi$ which can be easily calculated from the
saddle point conditions (\ref{spaconditions}). It is clearly seen from the
above expressions that the fluctuations fall off in the large area regime. As
far as the positivity of the mean square fluctuations(r.m.s. fluctuation to be
real) is concerned one needs to be more careful with the lower bound of
$\rhob$. One can show that there arise the following two cases :

\noindent 1)For $\sigmab<\frac{1}{3}\left(1+\frac{24}{\A}\right)$, the correct
lower bound for $\rhob$ is what has been obtained i.e. $1-3\sigmab\left(1-\frac{24}{\A}\right)$.

\noindent 2)For $\sigmab>\frac{1}{3}\left(1+\frac{24}{\A}\right)$, the correct
lower bound for $\rhob$ is given by $1-3\sigmab\left(1-\frac{16}{\A}\right)+\frac{8}{\A}$. 

\section{Discussion}
\label{DISC}

The first thing to say about this approach is that its origin is purely based
on quantum aspects of spacetime. During the build up of the formalism,
\textit{no classical metric} is used. The construction of the partition
function is purely based on the ideas and results of \textit{Loop Quantum
Gravity} e.g. the use of Chern-Simons states, the splitting up of the total
Hilbert space, etc. and also on the Hamiltonian formulation of spacetimes
admitting weakly isolated horizons.  The entropy correction also follows from
the quantum theory. The classical metrics come to the picture only to be
tested.

 In course of this heuristic statistical mechanical approach of stability
analysis of black holes, broadly two assumptions are made. In classical
Hamiltonian GR it is known that the total Hamiltonian (gravity + matter)
vanishes. So, it is very logical to consider that the quantum total
Hamiltonian operator annihilates the bulk states of quantum matter coupled
spacetime. A similar argument follows for the assumption of the quantum
constraint on the volume charge operator. These two assumptions may be
considered to be one due to their fundamental similarity and they ultimately
give rise to a single quantum constraint.  

In Section(\ref{TH}), a second
assumption is made regarding the eigenvalue spectrum of the energy of the
black hole. It is already mentioned (in a footnote) that this second
assumption is not a strong one. If one studies \cite{firstlaw} carefully, the
classical mass function associated with the horizon is stipulated to be a
function of horizon area and charge. Again, this horizon area and charge are
the functions of the local fields on the horizon. Proper quantization of the
classical horizon area and charge will obviously lead to a well defined
boundary Hamiltonian operator. The fact that there
exists a quantum boundary Hamiltonian operator which  acts on the boundary
Hilbert space of the black hole is an assumption, since the exact form of such
a Hamiltonian operator is still unknown.  But the fact that its eigenvalue
spectrum is a function of eigenvalue spectra of the area and charge operators
is most likely a valid assumption, as it is bound to happen if such a
boundary Hamiltonian operator exists. It follows from the classical analog -
the mass associated with the horizon must be a function of the
horizon area and charge for a consistent Hamiltonian evolution\cite{firstlaw}.

In ref. \cite{cm2} where a similar approach
has been taken, a particular functional form of the mass in terms of the area
and charge had been used on an ad hoc basis. Such an ad hoc assumption has
been shown here to be quite redundant. This, therefore, is a significant
strength of the paper, relative to the earlier assay. Thus, the statistical
mechanical approach adopted in this
paper, though similar, now stands on a far stronger ground than in the
previous version. 

This statistical mechanical approach gives us a new quantum correction to the
surface gravity arising from the loop quantum gravity corrections to the
microcanonical entropy. One can easily check its validity. Moreover, it predicts
local thermodynamic instability of the Reissner-Nordstrom black hole and local
thermodynamic stability of the AdS Reissner-Nordstrom black hole as was shown
classically in \cite{hp}. As far as the relative charge fluctuations are
concerned, in literature \cite{tapo} there are problems for AdS RN black holes
as it does not fall off for any condition. This problem has been solved in
this paper(Section\ref{VSPA}). Last but not the least, once more a word is
worth mentioning that this formulation \textit{does not} involve any classical
metric. The whole thing depends on quantum aspects of spacetime. One can
generalize this to study rotating and even charged-rotating black holes
because there is no use of symmetry in the theory. We look forward to give
those analyses in future.

There is a more crucial issue which can be of utmost importance to have a deeper understanding of black hole thermodynamics \cite{asen}. As far as the saddle-point
approximation is concerned, the Euclidean path integral approach does
look similar to our thermal holographic approach. But there is a
crucial difference between the two approaches. The path integral
approach, within the saddle point approximation, requires information
of the full black hole spacetime (bulk and
horizon) as given by the classical black hole metric solution chosen
to be the saddle point. Thus one needs global
information away from horizon. In the thermal holographic
approach which we adopt, one needs only {\it local information}
associated with the equilibrium isolated horizon geometry interpreted as an inner
null boundary. Thus, no detailed knowledge of the full classical
spacetime is needed. The mass of the isolated
horizon is an {\it unspecified} function of the area and charge, and
we never need to specify this function to derive our results, except
as fiducial checks [Subsection (\ref{CMTS})] appropriate to given classical
metrics. This insensitivity of our approach
to an explicit classical black hole metric is a key feature of our
work and can be taken to mean that our results are in a sense more
general than those computed from the Euclidean path integral.  

A further distinction is that, in contrast to the Euclidean path
integral approach, where quantum fluctuations around a {\it classical} metric
are considered, our saddle point is a {\it quantum} isolated horizon
whose quantum states and their (non-perturbative) dynamics are described by a quantum
Chern Simons theory. Consequently, the {\it equilibrium} entropy is
the microcanonical entropy computed in earlier work (ref. \cite{ack,kmplb,kmprl,abck}) based
on Loop Quantum Gravity, and already has an infinite series of
corrections (including those logarithmic in horizon area) beyond the
Bekenstein-Hawing area law, incorporating quantum spacetime fluctuations. In this paper, {\it
  additional thermal fluctuations} (and their physical effects) are considered, over and above the
quantum spacetime fluctuations already incorporated for the
equilibrium configuration. Quantum and thermal fluctuations are thus,
treated somewhat distinctly in our approach, and the result is an
interesting interplay between them. In the Euclidean path integral,
such distinctions are not as clear.  It might be of future interest to see better how these two somewhat
disparate approaches may be related.

\textbf{Acknowledgments :} We thank T. Sarkar for many illuminating
discussions during the course of this work
and also for bringing reference \cite{monteiro} to our attention. 

\end{document}